\PassOptionsToPackage{bookmarks=false}{hyperref}

\documentclass[sigconf]{acmart}

\copyrightyear{2020}
\acmYear{2020}
\setcopyright{acmcopyright}
\acmConference[MSR '20]{17th International Conference on Mining Software
Repositories}{October 5--6, 2020}{Seoul, Republic of Korea}
\acmBooktitle{17th International Conference on Mining Software Repositories (MSR
'20), October 5--6, 2020, Seoul, Republic of Korea}
\acmPrice{15.00}
\acmDOI{10.1145/3379597.3387451}
\acmISBN{978-1-4503-7517-7/20/05}

\usepackage[utf8]{inputenc}
\usepackage[english]{babel}
\usepackage{graphicx}
\usepackage{hyperref}
\usepackage{listings}
\usepackage{array}
\usepackage{amsmath}

\begin{document}

\title{Visualization of Methods Changeability Based on VCS Data}

\author{Sergey Svitkov}
\affiliation{Saint-Petersburg State University}
\email{svitkovsergey@gmail.com}

\author{Timofey Bryksin}
\affiliation{JetBrains Research\\
Saint-Petersburg State University}
\email{t.bryksin@spbu.ru}

\begin{abstract}
Software engineers have a wide variety of tools and techniques that can help them improve the quality of their code, but still, a lot of bugs remain undetected. In this paper we build on the idea that if a particular fragment of code is changed too often, it could be caused by some technical or architectural issues, therefore, this fragment requires additional attention from developers. Most teams nowadays use version control systems to track changes in their code and organize cooperation between developers. We propose to use data from version control systems to track the number of changes for each method in a project for a selected time period and display this information within the IDE's code editor. The paper describes such a tool called Topias built as a plugin for IntelliJ IDEA. Its source code is available at \url{https://github.com/JetBrains-Research/topias}. A demonstration video can be found at \url{https://www.youtube.com/watch?v=xsqc4gCTxfA}.

\end{abstract}

\maketitle

\section{Introduction}

Development of software products is a very complex process that often requires the joint work of multiple software developers. Various practices and techniques were introduced that help to keep the quality of code at an acceptable level, including static code analysis or thorough code reviews, but still, a lot of bugs remain undetected.

Several papers report a high correlation between how often bugs are found in certain code fragments and how often these fragments are changed over time~\cite{giger2012fine, herzig2013classifying, krishna2018connection}. A method or a function could change frequently due to some \textit{technical} (some hardcoded parameters should be moved to a configuration file), \textit{architectural} (this piece of code performs more than one task and should be split into several entities), or even \textit{external} (a particular business rule changing too often) issue. Wang et al.~\cite{wang2010detect} also report that if a bug is found in a code fragment, there is a high possibility that this fragment contains another bug. A successful bug fix might also take more than one attempt: for example, Yin et al.~\cite{yin2011fixes} report that up to 25\% of the bug fixes they examined were incorrect and required repeated changes. 

All this leads us to understand the value of raising developers attention towards frequently changed code fragments. If a developer is notified about such a piece of code, they could look at it more closely and possibly reveal and fix the issue behind it, for example, improve code structure or fix a bug. In order to be used on a regular basis, this tool should be integrated into a development environment that developers use in their work. Otherwise, the chances for it to be run are reduced significantly since developers usually are very reluctant to break the flow and switch to external tools from their IDE~\cite{johnson2001you}. 

This paper describes the implementation of a plugin for \mbox{IntelliJ} IDEA\footnote{IntelliJ IDEA, URL:~\url{https://www.jetbrains.com/opensource/idea/}} called Topias that collects data from a version control system's (VCS) change history for a given Java project, builds a change frequency model for each method, and shows it in the IDE's editor as read-only labels and bar charts placed next to the methods signatures. 

The remainder of this paper is structured as follows. Section~\ref{relatedwork} provides an overview of similar tools that try to draw developers attention to important parts of their projects. Section~\ref{overview} focuses on IntelliJ Platform's infrastructure and components needed to implement such a tool, examines tools for automatic refactoring detection and describes the \textit{RefactoringMiner} tool. Section~\ref{implementation} presents the proposed plugin's workflow and highlights the most important implementation details. Section~\ref{plugin} presents a visualization of the created tool and discusses its user interaction steps. 

\section{Related work}\label{relatedwork}   

This work was inspired by two existing tools: $code\_call\_lens$~\cite{janes2018code_call_lens} and $Azurite$~\cite{Yoon2013}.

The code\_call\_lens tool also tries to draw developers attention to important parts of their projects, but it is based on tracking functions' usage frequency. This information allows developers to estimate the impact of changing a particular function more accurately: if a function is called a lot and it breaks, the whole project could suffer.

 To gather such statistical data, a logging component is added to the target application. When the application is run, this component automatically catches each function call and sends this data to a server for processing. This tool is implemented as a plugin for Visual Studio Code and works for applications written in Python. It communicates with the server via REST API and displays the number of each function's calls next to its signature using read-only labels right in the IDE's editor.
 
 According to the authors, this data collection pipeline does not affect the overall performance of the application since all the logging is done in the background. The server processes the incoming data in-memory, which makes it possible to see the number of calls changing in nearly real-time as the application is running (the authors report that it takes about a second for the data to update). One of the known problems with code\_call\_lens is that function names are used as parts of their identifiers. If a function name is changed, all the collected data for it is lost.  

From our point of view, code\_call\_lens is an excellent example of a tool for detection and visual representation of the critical parts of a project. However, the issues with losing data when a function name is changed and more importantly the requirement of a standalone server greatly limit its use in practice. Inspired by this work, we decided to try a different approach to finding code fragments that require additional attention, while using visualization techniques similar to theirs. 

The idea to analyze code change in a project is not new. Yoon et al.~\cite{Yoon2013} present the Azurite tool that aims to reduce the effort to query the codebase: e.g., answer questions such as who introduced this piece of code, when, and why. Our goal is to target the previous step: attract developers’ attention to the possibly problematic parts of their system --- to raise the kind of questions that Azurite helps to answer. Another feature of Azurite is the visualization of the project’s timeline. While this is close to what we intend to do, their visualization is file-based, which gives a more coarse-grained overall view of the project. We believe that Topias complements Azurite very well and finds its place in between Azurite's features: e.g., a developer observes that a particular file is being changed too often (Azurite), opens this file and checks which methods were changed the most (Topias) and then tries to find the answer why (Azurite again).

\section{Overview}\label{overview}
In order to implement the proposed tool, the following steps should be taken:
\begin{itemize}
    \item read the commit history (each commit individually and all of them as a whole);
    \item store the processed commit history data internally to prevent recalculation on each IDE restart;
    \item build syntax trees for each file and each commit's content;
    \item detect method refactoring events in order to update the internal model correctly when a method is changed (for example, when a method is renamed);
    \item show the obtained change frequency data in the IDE's editor.
\end{itemize}

This section describes IntelliJ Platform's components and external libraries that help to solve these tasks.

\subsection{IntelliJ Platform}

\subsubsection{Program Structure Interface}
Program Structure Interface\footnote{IntelliJ Program Structure Interface, URL:~\url{https://www.jetbrains.org/intellij/sdk/docs/basics/architectural\_overview/psi.html}} (PSI) is the IntelliJ Platform's core. It provides a common application programming interface (API) for working with code in different programming languages. Each code fragment is represented by a PSI tree --- a syntax tree that contains all information about the code and provides a rich interface to work with it. There are different tree node types for different entities of a program: classes, methods, fields, comments, etc. This tree could also be built from a string object containing source code, which is useful when the original file is already deleted, but its revisions are still available in a VCS history. 

\subsubsection{Git4Idea}
The Git4Idea module\footnote{Git4idea, URL:~\url{https://github.com/JetBrains/intellij-community/tree/master/plugins/git4idea}} is a part of IntelliJ Platform that allows it to work with Git. It provides a rich object-oriented API for working with branches, changes history, revisions of files, etc. A single commit is represented by the \textit{GitCommit} class, storing the information about changes made in a commit, commit's hashcode, date and time, its author and a comment message. Each commit object stores two revisions of each changed file: before and after the commit. The revisions are stored as text strings, and we can build PSI trees from them using the IntelliJ Platform's API. 

\subsubsection{Editor}
Editor is an API for working with the text editor of IntelliJ Platform-based IDEs. It provides access to various features like working with currently opened file's text, modification of different visual components, events handling, etc. Visual elements of the editor are represented by its \textit{InlayModel} object, which has a wide variety of methods for adding, removing, and editing them. A wide range of visualizations is available, such as simple text labels, graphs, charts, or highlighting of the method's signature.

\subsection{RefactoringMiner}
As was mentioned in Section~\ref{relatedwork}, refactoring could cause issues with updating the collected data model. In order to avoid data loss when, for example, a method was renamed, cases like this should be deliberately tracked and processed. Several refactoring detection tools exist, such as RefFinder~\cite{kim2010ref}, JDEvAn~\cite{xing2008jdevan}, RefactoringCrawler~\cite{dig2005automatic}, or RefactoringMiner~\cite{tsantalis2018accurate}. The first three are implemented as Eclipse plugins, and the last one is a standalone library based on Eclipse's code analysis subsystem.

For our work, the RefactoringMiner tool has been chosen as it has the best reported accuracy metrics values among these tools (precision 0.97, recall 0.87) and is distributed as a library. It works with Java projects and is able to detect multiple types of refactoring on different levels. 

Despite RefactoringMiner being the best available tool that detects refactoring events within commits, it only supports programs written in Java and works with Git as a version control system. Another significant drawback is that it is built around Eclipse's code analysis framework, which in its turn brings a lot of these dependencies to our plugin. 

\section{Implementation}\label{implementation}

\subsection{Commit data processing pipeline}

\begin{figure*}[h]
        \centering
        \includegraphics[width=.8\textwidth]{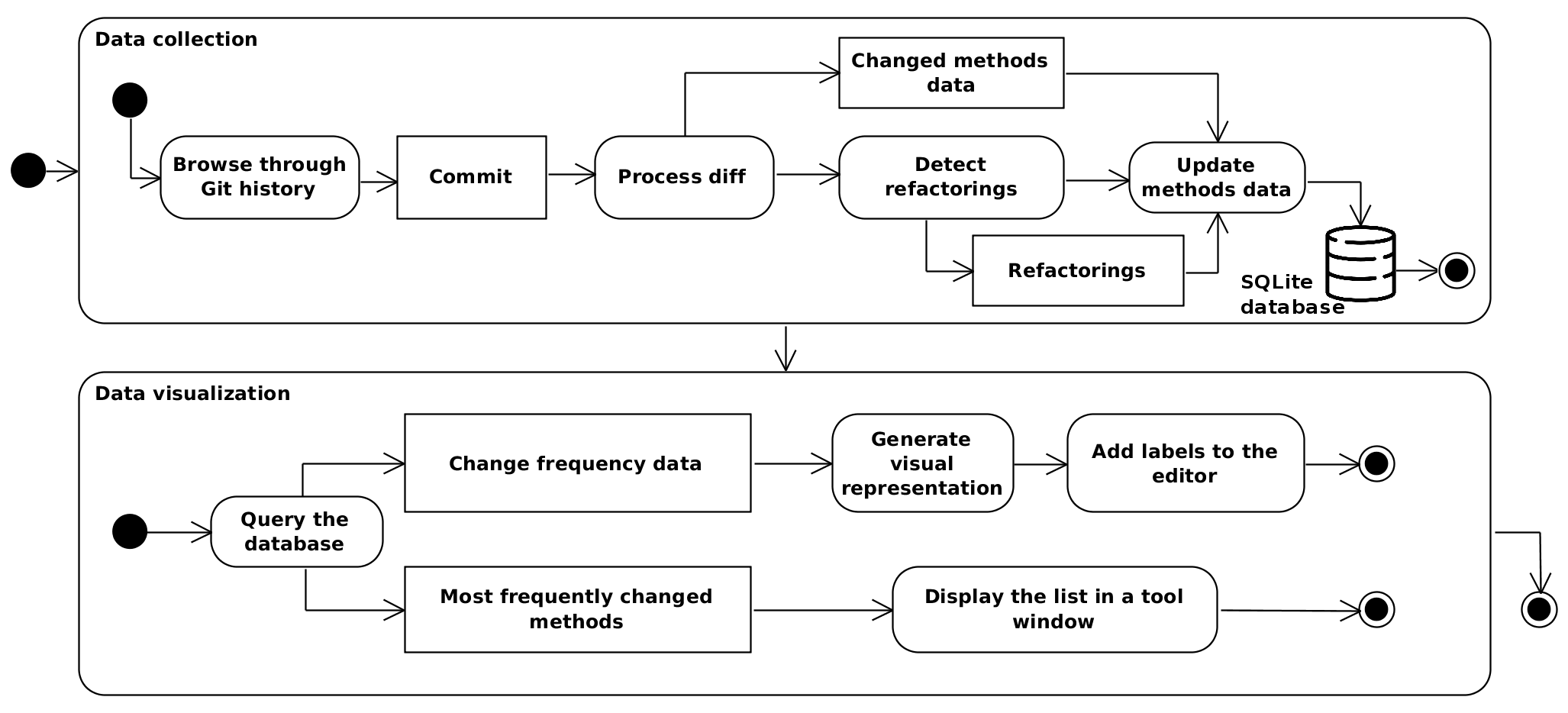}
        \caption{The pipeline of the proposed tool}
        \label{activity}
\end{figure*}

Figure~\ref{activity} presents the overall tool workflow. Commit data processing is triggered when either a project is opened in the IDE or a commit is made via the IDE's user interface. In the latter case, the tool proceeds with refactoring detection step. In the former case, if the project is new, all commits for a selected time period\footnote{Default time period is seven days, could be changed via the plugin's settings dialog.} are sequentially processed. Otherwise, each new commit is checked for having been already processed before by comparing its hash code against hash codes of previously processed commits. When the first unseen commit is found, the refactoring detection step begins.

\subsubsection{Refactorings detection}
Refactoring events are represented by objects wrapping RefactoringMiner's API. After the detection is finished, the handler returns a \textit{RefactoringData} object, holding information about affected methods before and after the refactoring. When all refactoring cases detected within one commit are processed, a collection of RefactoringData objects is available for further analysis.

\subsubsection{Analysis of commit changes}
The next step is the analysis of the commit changes. The plugin takes the \textit{before} and \textit{after} revisions of the file and builds PSI trees from their content in order to get information about methods in the changed file. Then it is used to update the plugin's current in-memory model: a collection of MethodInfo objects holding information about methods in this particular file. 

\subsubsection{Application of detected refactoring events}
If the obtained collection of the RefactoringData objects is not empty, the collection of MethodInfo objects is updated again according to the type of the detected refactoring:
\begin{itemize}
    \item Extract Method/Extract and Move Method refactoring: a MethodInfo object is created for the new method;
    \item Inline Method refactoring: a MethodInfo object for the inlined method is deleted, number of changes for a method, which this method was inlined into, is incremented;
    \item Rename Method refactoring: the old method name is replaced by the new one in the corresponding MethodInfo object;
    \item Move Method refactoring: the MethodInfo object is moved to a collection corresponding to the file which this method was moved to. The method's name in the MethodInfo object is updated since the method's full name is changed;
    \item Pull Up/Push Down Method refactoring: similar to the previous case, the MethodInfo object for pulled up/pushed down method is just moved to a new collection.
\end{itemize}

To prevent repeated browsing through the VCS history, the obtained in-memory model of the method changes is stored in a SQLite database. The aggregated statistical data for all methods is stored in a separate table, which is updated incrementally via SQLite's UPSERT queries when commits are processed. That way, change frequency data for each method is precalculated and is always available for visualization on demand.

\subsection{Visualization}
Visualization of the collected data is performed via separate visual blocks added to the editor's InlayModel. These elements are added and displayed when a new file is opened in the editor, as shown in Figure~\ref{activity}. The plugin's database is queried for all data available for this file, for each of the received MethodInfo objects a new visual element is created in the IntelliJ Platform's InlayModel.

\begin{figure}[h]
        \centering
        \includegraphics[width=\columnwidth]{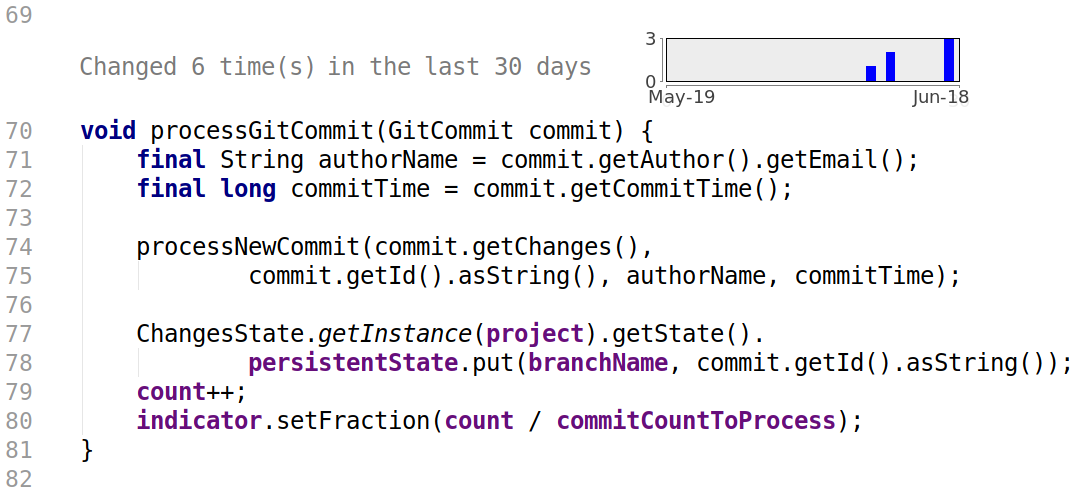}
        \caption{An example of methods change frequency visualization (located between lines 69 and 70)\label{visual}}
\end{figure}

Visual elements could be rendered in two ways (see Figure~\ref{visual}). The first representation is textual read-only labels displaying the number of method changes in the selected time period. The second representation is based on bar charts showing the number of changes by the day. Each renderer is responsible for proper visualization of its visual elements, including calculation of their widths, heights, and correct positions. For example, the label renderer uses current font settings of the IDE to perform such calculations.

These visualizations are inspired by the research of Harward et al.~\cite{Harward2010}, who approach the same task of attracting developers’ attention to interesting code fragments, but mostly rely on code coloring inside the editor. While they provide an impressive list of possible augmentations for different cases, we focus on a specific task of methods changeability. To raise developers’ awareness we display read-only labels and bar charts that blend in with the rest of the text editor and don’t distract from reading and writing the code itself. We considered going further, following Harward et al.’s approach and color lines of code within the methods according to their time of modification but in the end decided that it would be too distracting.

\section{The proposed plugin}\label{plugin}
The plugin is available at JetBrains Plugin Repository,\footnote{Topias at JetBrains Plugin Repository, URL: \url{https://plugins.jetbrains.com/plugin/12564-topias}} its source code is available on Github.\footnote{Topias on GitHub, URL: \url{https://github.com/JetBrains-Research/topias}}

When a developer opens a project in the IDE, the plugin checks if a version control system is used. If no VCS root directory is found, a warning message is shown, the plugin turns off and remains in this state until a new project is opened or a VCS root is set for the current project.

All data retrieval from a VCS history is done in the background. The full commit history for a selected time period is only browsed through once when a project is opened for the first time. After that, there is almost no impact on the overall IDE performance. When a new file from a current project is opened in the editor, the plugin queries its data model and builds visual elements for this file's methods. This is also done in a background thread, and therefore does not affect the editor performance, but it could take an extra half a second for the visual elements to appear, which might annoy some very impatient developers. 

In addition to the visual elements in the IDE's editor, the plugin also adds a new tool window showing a list of the top 10 most frequently changed methods within this project. By default the list is placed in the bottom IDEA’s tool panel. Clicking on a method in this list allows navigating to this method's declaration.

\section{Evaluation}
As a preliminary evaluation, we have asked five software engineers to install our plugin and to use it for a week, and conducted an interview with them afterward. All interviewed developers are working at the same company and work on banking systems written in Java. They have from 2 to 7 years of experience. We asked them to rate on the scale from 1 to 5 the tool’s UI (the average result we got is 4.6), the accuracy of the gathered data (average 4.2), the tool’s performance (average 4.6), and the overall usefulness of the tool (average 4.2). 

We have also asked if any bugs or architectural defects were found inside the methods that changed the most (yes or no question). In the following informal interviews with these developers, they mentioned finding issues such as immature class hierarchies (causing frequent change of inherited methods), unstable communication protocols between components and contracts between classes. The respondents who had not found any defects while using our tool noted that they found it useful (and interesting) to realize how often the code is actually changed. They agreed that the most stable parts of their software were changed less than others.

All of the interviewed developers reported that the visualization was smooth and did not have any negative effect on the performance of their IDEs.

As future work, we plan to run an extensive study based on the developed tool to see how the developers' awareness of frequently changed code fragments affects the overall code quality.

\section*{Conclusion}\label{conclusion}
In this paper, we present Topias --- a tool for the visualization of methods change frequency according to the version control system data. It was implemented as a plugin for IntelliJ IDEA supporting Java programming language and Git VCS. These limitations come from the RefactoringMiner tool that we use to detect applied refactorings. If there were similar tools for other languages or VCS, the plugin could easily be extended to support other IDEs built on IntelliJ Platform.

\nocite{*}
\def\BibTeX{BibTeX}
\bibliographystyle{ACM-Reference-Format}
\bibliography{paper} 

\end{document}